# Quantifying Loss Aversion in Cyber Adversaries via LLM Analysis

Soham Hans, Nikolos Gurney
USC Institute for Creative Technologies
Playa Vista, CA
sohamhan@usc.edu, gurney@ict.usc.edu

Stacy Marsella, Sofia Hirschmann
Northeastern University
Boston, MA
s.marsella@northeastern.edu,
hirschmann.so@northeastern.edu

## ABSTRACT

Understanding and quantifying human cognitive biases from empirical data has long posed a formidable challenge, particularly in cybersecurity, where defending against unknown adversaries is paramount. Traditional cyber defense strategies have largely focused on fortification, while some approaches attempt to anticipate attacker strategies by mapping them to cognitive vulnerabilities, yet they fall short in dynamically interpreting attacks in progress. In recognition of this gap, IARPA's ReSCIND program seeks to infer, defend against, and even exploit attacker cognitive traits. In this paper, we present a novel methodology that leverages large language models (LLMs) to extract quantifiable insights into the cognitive bias of loss aversion from hacker behavior. Our data are collected from an experiment in which hackers were recruited to attack a controlled demonstration network. We process the hacker-generated notes using LLMs using it to segment the various actions and correlate the actions to predefined persistence mechanisms used by hackers. By correlating the implementation of these mechanisms with various operational triggers, our analysis provides new insights into how loss aversion manifests in hacker decision-making. The results demonstrate that LLMs can effectively dissect and interpret nuanced behavioral patterns, thereby offering a transformative approach to enhancing cyber defense strategies through real-time, behavior-based analysis.

## ABOUT THE AUTHORS

**Soham Hans** is a Research Project Specialist with the Human-Inspired Adaptive Teaming Systems Group at the University of Southern California's Institute for Creative Technologies. His work involves leveraging generative language models for cognitive analysis, particularly in cybersecurity and adversarial reasoning. He also applies multimodal generative AI to support scenario-based environments, with an emphasis on enhancing reinforcement learning through realistic, cognitively grounded simulations.

**Dr. Nikolos Gurney** is a Research Associate at the USC Institute for Creative Technologies. He earned his PhD in Behavioral Decision Research at Carnegie Mellon University. Dr. Gurney's research focuses on aligning models of human and AI decision making, cognition, and teaming. He is the interim research lead of the Institute for Creative Technologies Social Simulation lab, heading DoD-funded projects in cyber security, human-machine integration, and AI social reasoning.

**Dr. Stacy C. Marsella** is a Professor at Northeastern University in the Khoury College of Computer Sciences with a joint appointment in the Department of Psychology. His research is in the computational modeling of human cognition, emotion, and social behavior, both as a basic research method in the study of human behavior as well as for use in a range of applications. His work has been applied to the modeling of human behavior for large scale social simulations, realization of effective human-AI teamwork as well as the design of intelligent social agents, software entities that can interact with humans using mulitmodal behavior.

**Sofia Hirschmann** is a doctoral student in the Khoury College of Computer Sciences at Northeastern University, advised by Stacy Marsella. Before joining Khoury College in 2023, she completed a bachelor's in science at Simmons University in Cognitive Psychology and Data Science. She is interested in artificial intelligence, with her research investigating methods to improve AI's ability to model humans in both collaborative and adversarial domains. Within the Cognitive Embodied Social Agents Research (CESAR) Lab at Northeastern, she specifically focuses on methods of computationally modeling cognition.

# Quantifying Loss Aversion in Cyber Adversaries via LLM Analysis

Soham Hans, Nikolos Gurney
USC Institute for Creative Technologies
Playa Vista, CA
sohamhan@usc.edu, gurney@ict.usc.edu

Stacy Marsella, Sofia Hirschmann
Northeastern University
Boston, MA
s.marsella@northeastern.edu,
hirschmann.so@northeastern.edu

## INTRODUCTION

Defending against unknown, adaptive adversaries remains one of the most persistent and complex challenges in cybersecurity. While significant advances have been made in fortifying digital infrastructure, the ever-evolving strategies of sophisticated attackers continue to outpace static defense systems. Traditional cybersecurity approaches tend to rely heavily on perimeter fortification, signature-based detection, and reactive patching—tools which, although important, often prove insufficient when confronting novel or adaptive adversary behaviors.

In contrast, a growing body of research has sought to interpret attacker behavior more dynamically, often by mapping actions to known cognitive vulnerabilities or behavioral tendencies. These approaches aim to anticipate attacker strategies through psychological modeling, but they generally stop short of real-time interpretation or exploitation of these traits during an active intrusion. They also suffer from limitations in terms of observability and inference—understanding what an attacker is doing and why, while they are doing it, is not a solved problem.

Recognizing this gap, the Intelligence Advanced Research Projects Activity (IARPA) launched the ReSCIND (Reimagining Security with Cyberpsychology-Informed Network Defenses) program to rebalance the asymmetry of cyber conflict by exploiting attacker cognitive traits. ReSCIND leverages insights from cyberpsychology to develop defenses that detect, induce, and manipulate human vulnerabilities during cyber operations.

A key motivation of this work is to enable real-time interpretation of attacker behavior by analyzing the decision-making processes that unfold during an intrusion. Among various cognitive biases, one particularly relevant to adversary decision-making is **loss aversion**—a well-documented phenomenon in behavioral economics where individuals exhibit a tendency to prefer avoiding losses over acquiring equivalent gains. In cyber operations, loss aversion may manifest as a preference to preserve access to a compromised system or to maintain persistence mechanisms even when risks of detection rise.

This paper presents a novel methodology for identifying evidence of loss aversion in cyber adversaries by leveraging large language models (LLMs) to analyze behavioral data. We focus specifically on operational notes (OPNOTES) written by hackers participating in an incentivized red-team experiment against a controlled demonstration network. These notes, written during the course of the attack, provide a rare window into the intruders' thought processes and decision sequences.

Our approach involves segmenting the hacker notes into discrete actions using LLMs, then correlating these actions with known categories of persistence mechanisms. By examining the timing, frequency, and conditions under which these mechanisms are implemented, we extract patterns that signal loss-averse behavior. This proof-of-concept study offers compelling evidence that LLMs can serve as powerful tools for real-time interpretation of adversary behavior—dissecting and quantifying cognitive patterns that were previously difficult to observe systematically.

In doing so, we contribute to a new paradigm of cyber defense—one that augments static defenses with adaptive, behavior-aware systems capable of anticipating and countering threats based on adversary cognition. Our results demonstrate the feasibility of using LLMs not only as linguistic processors but also as behavioral interpreters, bridging the gap between observable attacker activity and the underlying decision-making biases that drive it.

## RELATED WORK

**Intrusion Detection and the MITRE ATT&CK Framework**

The MITRE ATT&CK framework has become a foundational resource for modeling adversary tactics, techniques, and procedures (TTPs) in intrusion detection and threat classification systems. It offers a structured ontology for interpreting cyberattacks beyond signature-based alerts, and recent work has leveraged it for labeling malware behaviors, enriching threat intelligence, and enhancing rule-based intrusion detection systems. Al-Sada et al. (2024) and Roy et al. (2023) survey the breadth of ATT&CK usage across academic and industrial settings, noting its growing role in aligning observed activity with behavioral models. Oosthoek and Doerr (2019) demonstrated the utility of ATT&CK in categorizing post-compromise malware techniques, revealing shifts in attacker tactics over time.

**Cognitive Vulnerabilities in Cyber Adversaries**

While much of the cognitive security literature has focused on the limitations of human defenders, recent work has begun to examine the psychological biases and heuristics exhibited by attackers themselves. Studies from controlled red-team exercises, such as the Ferguson-Walter et al. (2018), have shown that adversaries can exhibit the default effect, sunk-cost fallacy, and confirmation bias during engagements. Hitaj et al. (2025) extended this insight by designing cognitive traps that exploit attacker heuristics—for instance, presenting decoy targets that align with an adversary's representativeness heuristic, thereby increasing the likelihood of engagement. These findings support a growing cognitive cybersecurity paradigm, where defensive strategies are informed by adversary psychology. In particular, the manifestation of *loss aversion*—a bias toward avoiding losses even at greater risk—has been observed in attacker persistence strategies and decision fatigue during prolonged engagements.

**LLMs for Human Behavior Modeling and Annotation**

Large language models (LLMs) have shown increasing promise in modeling human psychological traits and decision patterns from natural language. Peters and Matz (2024) found that LLMs such as GPT-3.5 and GPT-4 could infer Big Five personality traits from social media posts with correlations comparable to traditional psychometric models. Bunt et al. (2025) demonstrated that GPT-4 could function as a behavioral annotator, reliably classifying text data across diverse domains such as empathy labeling and sociolinguistic repair, often matching human-level performance when carefully prompted and validated. Similar results were reported by Gilardi et al. (2023) and Törnberg (2024), where LLMs outperformed crowd workers in subjective annotation tasks like political stance, emotion, or framing. These capabilities point to LLMs as scalable tools for behavioral inference and psychological labeling, supporting their use in extracting cognitive patterns from hacker-authored content.

**LLMs in Cybersecurity**

LLMs have been applied across cybersecurity domains, from threat intelligence summarization to vulnerability detection and adversary emulation. Daniel et al. (2025) showed that LLMs can effectively tag Snort rules and enrich threat reports by linking low-level observables to ATT&CK techniques or threat actor profiles. Zhou et al. (2024) demonstrated that models like GPT-4 and CodeLLaMA can detect vulnerabilities in C/C++ and Java with increasing accuracy as context size grows, rivaling traditional static analysis tools. Additionally, Wang et al. (2024) incorporated chain-of-thought reasoning into code assessment, improving the model's interpretability and detection of subtle flaws. These studies underscore LLMs' potential not only to assist defenders, but also to interpret attacker behavior and strategy based on textual artifacts.

**Research under IARPA ReSCIND and related government initiatives**

IARPA ReSCIND, *Reimagining Security With Cyberpsychology-Informed Network Defenses I*, will add new tools to the cyber-defender's toolkit. Rather than focusing on the hard skills of attackers or the fortification of network attached resources, ReSCIND seeks to leverage known cognitive vulnerabilities and human limitations to impede the progress of malicious agents. The core of this approach is identifying ways to manipulate the humans behind an attack. This means not only understanding how an attack progresses, but also the decision-making and related behaviors that take palace across the various stages of an attack.

**Psychological background on loss aversion and its operational relevance**

Loss aversion describes the tendency of people to over-weight losses relative to equivalent gains (Kahneman, D., & Tversky, A. (2013)). Concretely, this means that the experience of losing something already in one's possession, for example, tickets to a concert, is accompanied by a degree of emotional pain greater than the degree of elation that accompanies its inverse, for example, finding tickets to the same concert. ReSCIND hypothesizes that cyber attackers experience such emotions during attacks, that it is possible to detect when they are experiencing them, and that a well-informed defensive system can leverage them.

**RESEARCH METHODS**

**Experimental Setup**

Our analysis is grounded in data collected from *Operation 418*, a controlled academic cybersecurity experiment designed to examine hacker decision-making in a realistic, adversarial setting. Participants in the study—trained cybersecurity professionals acting as penetration testers—were tasked with compromising a simulated corporate network over a two-day engagement. The environment, modeled after a mid-sized enterprise, included segmented subnets, domain controllers, and security systems such as intrusion detection and incident response protocols.

Participants operated from isolated virtual machines under constrained conditions that limited external resources, simulating realistic operational restrictions. The data used in this study were collected independently as part of the IARPA ReSCIND program. The experiment followed a four-stage progression that reflected common phases of offensive operations: reconnaissance, privilege escalation, lateral movement, and data exfiltration. At each stage, participants were given increasingly detailed intelligence and broader access, allowing researchers to observe shifts in attacker strategies under evolving objectives and constraints.

Crucially, participants were required to document their activities through detailed, timestamped operational notes (OPNOTES). These notes formed the primary data source for our study, capturing decision-making processes, tool usage, and justification for actions taken. While self-report data introduces potential biases, the structured reporting protocol and real-time nature of documentation enhance its reliability. Additionally, researchers introduced dynamic updates (e.g., simulated maintenance alerts) to assess how participants adapted to unfolding information, providing insight into in-the-moment cognitive responses.

**Cognitive Bias Hypothesis**

This study investigates the presence of loss aversion in cyber adversary behavior during offensive operations. Loss aversion, a well-established cognitive bias, describes the human tendency to weigh potential losses more heavily than equivalent gains. In operational contexts, this often manifests as excessive efforts to preserve existing resources, even when doing so may not be rational from a purely utility-maximizing perspective.

To operationalize this concept within our experiment, we focus on the use of *persistence techniques* as defined by the MITRE ATT&CK framework. Persistence techniques allow an attacker to maintain access to a compromised system over time. Our hypothesis is that an adversary exhibiting loss aversion will show a heightened focus on maintaining access through the use of persistence mechanisms.

Our central hypothesis is that attackers exhibiting loss-averse behavior will allocate more attention and resources to implementing persistence mechanisms—even when not strictly required to accomplish their stated objectives. Such behavior may manifest in several ways, including increased frequency of persistence techniques, a broader variety of mechanisms employed, or prolonged time spent preparing and deploying such measures.

Our analysis emphasizes the breadth and depth of persistence behaviors across participants. By examining the number of distinct persistence techniques used, their temporal distribution throughout the operation, and the complexity of their implementation, we assess whether attacker decision-making reflects an elevated concern for retaining system access. This behavioral pattern would provide empirical support for the role of loss aversion as a cognitive driver in adversarial cyber operations.

**Data Processing & Annotation Pipeline**

To analyze behavioral patterns from participant-authored operational notes (OPNOTES), we developed a multi-stage annotation pipeline centered around large language model (LLM) reasoning. The objective of the pipeline was to transform free-form textual logs into structured sequences of attacker actions, enabling downstream analysis of cognitive traits and technical strategies.

We employed OpenAI's GPT-4o (2024) to parse each participant's notes and extract a sequence of discrete actions, each defined by a start and end point within the timeline of the exercise. For each action, the model generated a concise description of the task being performed. This segmentation allowed us to reconstruct the operational flow of each attacker's activity during the engagement.

Following segmentation, each action was analyzed in the context of its surrounding actions to determine whether it reflected a *persistence technique*, as defined by the MITRE ATT&CK framework. When relevant, the action was mapped to the closest corresponding MITRE Persistence sub-technique. Importantly, each classification was accompanied by explicit reasoning steps, prompting the LLM to justify its decisions prior to issuing final labels. This approach enhanced interpretability and allowed researchers to verify model outputs during manual review.

Rather than fine-tuning a model—an approach infeasible given the limited size of our dataset—we designed the LLM interaction process to mirror a stepwise analytical workflow. The modular structure defined above allowed us to scale the analysis across a large volume of textual data while maintaining transparency and consistency. By explicitly structuring the process into logical steps, we enabled more accurate assessments and facilitated post hoc verification of model outputs.

Through this LLM-based annotation pipeline, we generated a structured representation of attacker behavior that enables correlation between persistence-related decisions and potential cognitive triggers, forming the basis for our subsequent loss aversion analysis.

**Analysis Approach**

To assess the presence of loss-averse behavior in attacker decision-making, we adopt a correlational approach that links the observed use of persistence techniques to situational risk triggers introduced during the cyber operation. Our analysis focuses on how attackers' propensity to loss aversion affects their actions throughout the experiment.

Because direct measurement of cognitive bias was not feasible in the experimental setting, we draw on two established psychological instruments as conceptual surrogates: the *General Risk Propensity Scale (GRiPS)* and components of the *Adult Decision-Making Competence (ADMC)* assessment. GRiPS quantifies individual tendencies toward general risk-taking behavior across everyday contexts, with items such as "Taking risks makes life more fun" or "I commonly make risky decisions." In contrast, ADMC focuses on cognitive traits related to decision performance, including *Resistance to Framing*, which captures susceptibility to how problems are described in terms of gains or losses—a central mechanism in loss aversion.

In our analysis, we operationalize loss aversion by examining the frequency and nature of persistence techniques employed during cyber operations. Using the annotated action sequences extracted from participant notes, we map each persistence-related action to corresponding MITRE techniques and count their occurrences per participant. These counts serve as behavioral proxies for loss-averse tendencies, based on the premise that repeated efforts to maintain system access reflect a preference for avoiding losses over seeking gains.

To explore the relationship between these behaviors and individual cognitive traits, we perform both correlation analysis and multivariate regression modeling using psychometric measures described above. These models assess whether higher levels of risk aversion or susceptibility to framing effects are associated with increased use of persistence tactics. By aligning behavioral data with psychological constructs through interpretable statistical models, we aim to uncover measurable links between attacker cognition and technical decision-making.

**RESULTS AND FINDINGS**

**Summary Statistics and Behavior Distribution**

A total of 17 participants completed operational notes that met the requirements for analysis. Participants were also divided into two groups—Expert (n = 9) and Open (n = 8) divisions—based on their responses to a background questionnaire assessing cybersecurity experience.

Using our annotation pipeline, we extracted structured action sequences and identified associated MITRE ATT&CK persistence techniques for each participant.

On average, each participant employed approximately 13.88 persistence-related techniques over the course of the experiment (median: 13; range: 6–27). Across all participants, the pipeline identified a total of 14 unique MITRE persistence techniques, distributed across 56 unique subtechnique instances.

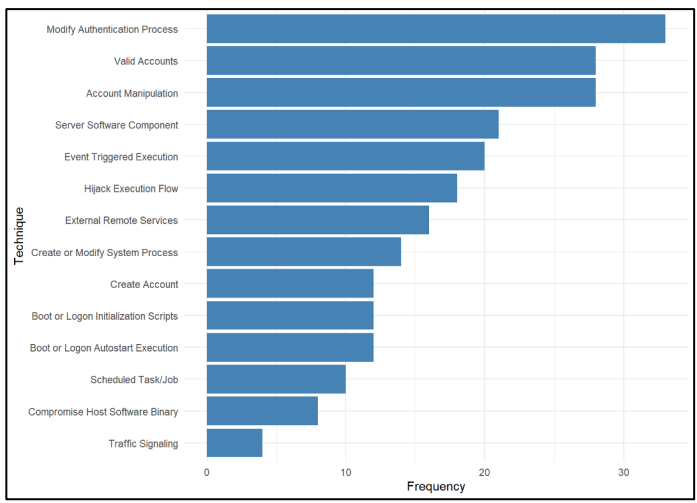

**Figure 1. Most Frequently used MITRE ATT&CK Techniques**

The three most frequently observed persistence techniques were:

1. **Modify Authentication Process** – 33 occurrences (**14.0%** of total)
2. **Account Manipulation** – 28 occurrences (**11.9%**)
3. **Valid Accounts** – 28 occurrences (**11.9%**)

These techniques indicate a strong emphasis on credential-based and access maintenance strategies among participants, consistent with the hypothesis that attackers prioritized actions aimed at sustaining existing system access.

The distribution of technique usage is visualized in Figure 1, which displays the relative frequency of each MITRE persistence technique observed across all annotated actions.

**Correlation between cognitive traits and technical attack patterns**

To explore the influence of cognitive traits on adversary decision-making, we conducted both correlation analysis and multivariate regression modeling, focusing on the relationship between psychometric scores and the use of MITRE persistence techniques during offensive cyber operations.

To investigate the impact of cognitive traits on adversary behavior, we analyzed how psychometric profiles related to the use of MITRE-defined persistence techniques during cyber operations. This included both Pearson correlation analysis and multivariate regression modeling.

We examined three psychometric measures: GRiPS, and two subscales from the ADMC inventory—Resistance to Framing for gains (RC1) and Resistance to Framing for losses (RC2)—which assess susceptibility to cognitive framing effects in decision-making.

**Correlation Analysis**
We first computed Pearson correlations between the total number of persistence techniques used by each participant and three psychometric measures. As shown in Table 1, GRiPS exhibited a moderate negative correlation with persistence usage (r = –0.43, p = 0.065), suggesting that participants with lower general risk-taking propensity—a surrogate for higher loss aversion—tended to employ more persistence techniques. While this relationship did not reach conventional statistical significance, it supports the hypothesis that risk-averse individuals exhibit behaviors consistent with loss aversion by prioritizing access retention.

In contrast, ADMC RC1 and RC2 scores were weakly and non-significantly correlated with persistence usage (r = –0.09 and –0.20, respectively), providing limited support for framing-related decision vulnerabilities as explanatory factors in this context.

**Table 1. Pearson Correlation between persistence technique usage and psychometric measures**

|  | CORRELATION | P_VALUE | CI_LOWER | CI_UPPER |
|---|---|---|---|---|
| **LA_GRiPS** | -0.43151994 | 0.065070683 | -0.74057605 | 0.0282206 |
| **LA_ADMC_RC1** | -0.09360824 | 0.703079390 | -0.52547550 | 0.3766138 |
| **LA_ADMC_RC2** | -0.19870651 | 0.414774546 | -0.59886558 | 0.2808508 |

**Regression Modeling**
To complement the correlation analysis, we constructed a multiple linear regression model to assess how cognitive traits predict adversarial use of persistence techniques. The dependent variable was the total number of MITRE-defined persistence techniques used by each participant. Independent variables included GRiPS, ADMC RC1, ADMC RC2 inventory, and the participant division (Open vs. Expert). **Table 2** summarizes the model coefficients.

**Table 2: Multiple Linear Regression Predicting Persistence Technique Count**

| Predictor | Estimate | Std. Error | t-value | p-value | Significance |
|---|---|---|---|---|---|
| **(Intercept)** | 25.021 | 6.175 | 4.052 | 0.001 | Highly Significant |
| **ADMC RC1 Score** | 3.869 | 2.345 | 1.650 | 0.121 | Not Significant |
| **ADMC RC2 Score** | –2.454 | 2.230 | –1.100 | 0.290 | Not Significant |
| **GRiPS Score** | –4.421 | 2.008 | –2.202 | 0.045 | Statistically Significant |
| **Division (Open)** | –4.775 | 2.791 | –1.711 | 0.109 | Marginally Significant |

The model revealed a statistically significant negative relationship between GRiPS scores and persistence usage ($\beta = -4.42$, $p = 0.045$), indicating that participants with lower general risk propensity employed more persistence techniques—consistent with behavior predicted by loss aversion. Figure 2(a) visualizes this relationship, showing a downward trend in persistence use with increasing GRiPS scores, especially among Open division participants

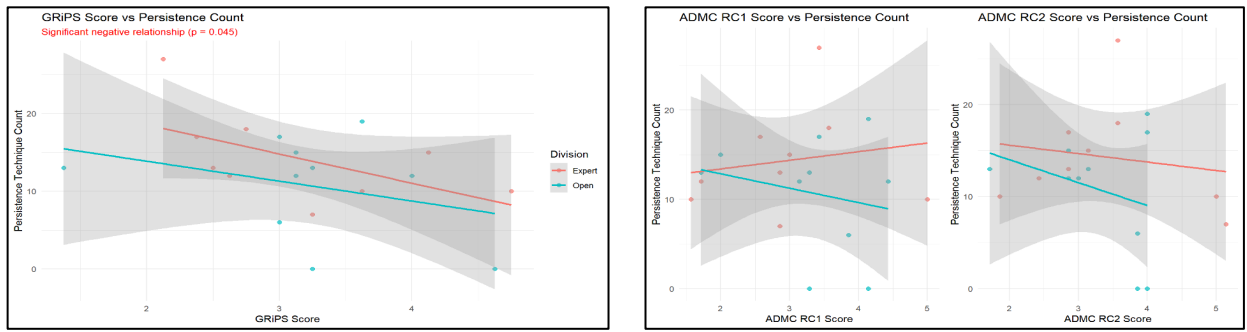

**Figure 2(a) (left): Scatterplot and linear fit showing the negative correlation between GRiPS scores and persistence technique usage. Figure 2(b) (right): Scatterplots showing the relationship between ADMC RC1 and RC2 scores with persistence usage across both divisions.**

The two ADMC subscales, RC1 (gain framing) and RC2 (loss framing), did not significantly predict persistence behavior ($p = 0.121$ and $p = 0.290$, respectively), though their raw associations are visualized in Figure 2(b). While the regression coefficients show opposing directions (positive for RC1 and negative for RC2), neither relationship achieved statistical significance, and their confidence intervals were wide, as reflected in Figure 3.

Division membership also showed a marginal effect ($\beta = -4.78, p = 0.109$), with Open division participants tending to use fewer persistence techniques than Expert participants (Figure 4). However, this relationship did not reach statistical significance at the conventional 0.05 threshold.

Model fit statistics showed that the regression explained approximately 37.6% of the variance in persistence technique usage ($R^2 = 0.376$), though the adjusted $R^2$ dropped to 0.198, reflecting reduced explanatory power after accounting for model complexity and sample size ($n = 17$). The overall model did not achieve statistical significance ($F(4,14) = 2.11, p = 0.133$), underscoring the need for replication with larger datasets.

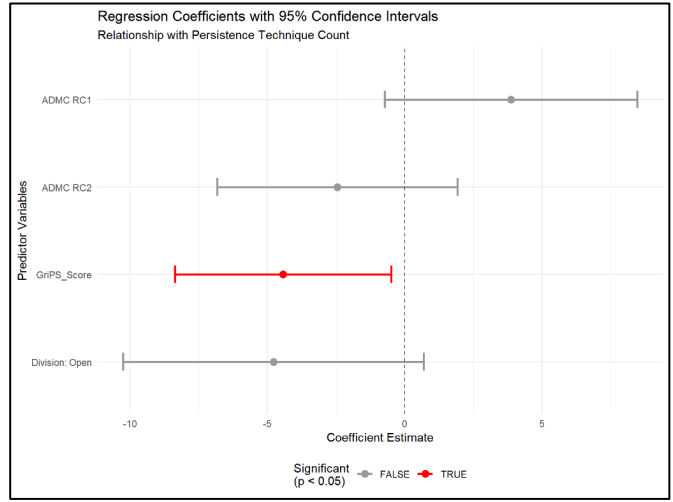

**Figure 3: Coefficient plot with 95% confidence intervals for all model predictors**

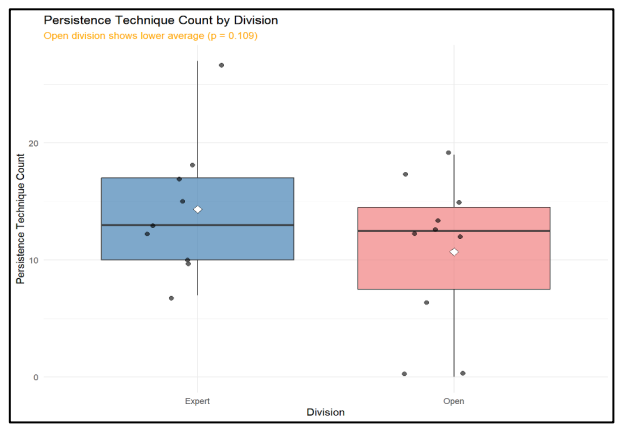

**Figure 4: Boxplot showing persistence technique counts by participant division (Open vs. Expert).**

Together, these findings suggest that general risk propensity (as measured by GRiPS), rather than framing-based vulnerabilities, plays a more consistent role in shaping persistence behaviors during offensive cyber operations.

**FUTURE WORK AND RESEARCH DIRECTIONS**

This study demonstrates the feasibility of using large language models (LLMs) to extract cognitively relevant features from human-authored cyberattack data. Building upon these initial findings, several promising avenues for future work emerge.

**Scaling to additional cognitive biases**

While this work focused on loss aversion, the broader spectrum of cognitive vulnerabilities—such as overconfidence,

confirmation bias, and sunk-cost fallacy—offers fertile ground for further investigation. Systematic annotation pipelines can be adapted to detect behavioral markers of these biases in attacker decision-making.

**Application to machine-generated data sources**

Our method currently relies on human-authored operational notes. Future iterations can extend this approach to machine-generated sources such as Suricata (OISF, 2024) or Zeek (Zeek Project, 2024) logs, applying LLMs to segment and interpret attacker behavior at scale without requiring explicit self-reports. This could bridge the gap between human-centric analysis and automated cyber monitoring.

**Real-time deployment for proactive defense**

An exciting application involves integrating these models into active defense platforms. By analyzing attacker behavior in near-real time, defenders could dynamically adjust controls based on inferred intent or cognitive patterns, transitioning from reactive security to anticipatory defense.

**Diversifying the dataset and adversary profiles**

The current study is limited in participant diversity. Future datasets should include broader demographics, skill levels, and cultural backgrounds to build more generalizable models of attacker cognition. Additionally, synthetic adversary profiles could be generated and tested in simulation environments.

These directions collectively aim to develop more cognitively-aware cybersecurity systems, capable of understanding and anticipating adversarial behavior at both strategic and operational levels.

## CONCLUSION

This work introduces a novel approach to interpreting human attacker behavior through large language models. By analyzing structured operational notes from cybersecurity exercises, we demonstrate that LLMs can reliably extract temporal action sequences, identify MITRE persistence techniques, and associate these actions with cognitive traits such as loss aversion.

Our findings provide empirical evidence that behavioral indicators of cognitive biases or tendencies—particularly risk aversion—can be inferred from attacker decisions in realistic red team scenarios. This not only advances the scientific understanding of adversarial reasoning but also offers practical utility for cyber defense planners seeking to anticipate and counter attacker strategies.

Ultimately, this study highlights the promise of LLMs as tools for cognitive analysis in cybersecurity, laying a foundation for next-generation defensive systems that leverage behavioral insight to stay ahead of adaptive threats.

## ACKNOWLEDGEMENTS

This research is based upon work supported in part by the Office of the Director of National Intelligence (ODNI), Intelligence Advanced Research Projects Activity (IARPA) under Reimagining Security with Cyberpsychology-Informed Network Defenses (ReSCIND) program contract N66001-24-C-4504. The views and conclusions contained herein are those of the authors and should not be interpreted as necessarily representing the official policies, either expressed or implied, of ODNI, IARPA, or the U.S. Government. The U.S. Government is authorized to reproduce and distribute reprints for governmental purposes notwithstanding any copyright annotation therein.